\documentclass[twocolumn,eqsecnum,floatfix,aps,prc]{revtex4}
\usepackage[dvips]{graphicx}
\def\rhovec{\mbox{\boldmath $\rho$}}
\def\fun#1#2{\lower3.6pt\vbox{\baselineskip0pt\lineskip.9pt
  \ialign{$\mathsurround=0pt#1\hfil##\hfil$\crcr#2\crcr\sim\crcr}}}
\usepackage{latexsym,amsbsy,amsmath}
\renewcommand{\vec}[1]{\boldsymbol{#1}}
\newcommand{\Jpsi}{{J \hspace{-2pt}/  \hspace{-2pt}\psi}}

\begin{document}

%\parindent=10pt
%%%%%%%%%%%%%%%%%%%%%%%%%%%%%%%%%%%%%%%%%%%%%%%%%%%%%%%%%%%%%%%%
\title{
Quark model  estimate of hidden-charm pentaquark resonances}

\author{Emiko Hiyama}
\email{hiyama@riken.jp}
\affiliation{Department of Physics, Kyushu University, Fukuoka, Japan,
819-0395}
\affiliation{Nishina Center for Accelerator-Based Science, RIKEN, Wako,
351-0198,  Japan}
\affiliation{Advanced Science Research Center, Japan Atomic Energy Agency, Tokai, Ibaraki, 319-1195 Japan}
\affiliation{Research Center for Nuclear Physics, Osaka University, Ibaraki, Osaka 567-0047, Japan}

\author{Atsushi Hosaka}
\email{hosaka@rcnp.osaka-u.ac.jp}
\affiliation{Research Center for Nuclear Physics, Osaka University, Ibaraki, Osaka 567-0047, Japan}
\affiliation{Advanced Science Research Center, Japan Atomic Energy Agency, Tokai, Ibaraki, 319-1195 Japan}
\author{Makoto Oka}
\email{ oka@th.phys.titech.ac.jp}
\affiliation{Department of Physics, H-27, Tokyo Institute of Technology, Meguro, Tokyo 152-8551, Japan}
\affiliation{Advanced Science Research Center, Japan Atomic Energy Agency, Tokai, Ibaraki, 319-1195, Japan}
\author{Jean-Marc~Richard}
\email{j-m.richard@ipnl.in2p3.fr}
\affiliation{Universit\'e de Lyon, Institut de Physique Nucl\'eaire de Lyon,
IN2P3-CNRS--UCBL,\\
4 rue Enrico Fermi, 69622  Villeurbanne, France}

\date{\today}
%
%%%%%%%%%%%%%%%%%%%%%%%%%%%%
\begin{abstract}
A quark model, 
which reproduces the ground-state mesons and baryons, i.e., the threshold energies,
 is applied to 
the $qqqc\bar c$ configurations, where $q$ is a light quark and~$c$ the charmed quark. 
In the calculation, several open channels are explicitly included  such as
$\Jpsi +N$, $\eta_c+N$, $\Lambda_c +D$, etc.
To distinguish genuine resonances and estimate their width,
we employ Gaussian Expansion Method supplemented by the real scaling
method (stabilization).    
 No resonance is found at the energies of the   $P_c(4380)$ and $P_c(4450)$ pentaquarks.
On the other hand, there is a sharp resonant state at 4690\,MeV with $J=1/2^-$
state and another one at 4920\,MeV with $J=3/2^-$ state,  which have a compact structure.
\end{abstract}

\maketitle

\section{Introduction}\label{se:intro}
During the past decades, several experimental candidates have been proposed for hadrons beyond the ordinary quark-antiquark and three-quark structures: baryonium, supernumerary scalar mesons, light pentaquarks, etc. However, most of them have not been confirmed in experiments with high statistics and better resolution, or their interpretation as multiquarks has a little faded away. A new era has begun with the discovery of the $X(3872)$ in 2003~\cite{Choi:2003ue} and several other $XYZ$ states with hidden charm. Another striking result is the observation of the $P_c$ pentaquarks by the LHCb collaboration \cite{Aaij:2015tga}. For a review, see for instance~\cite{Hosaka:2016pey,Chen:2016qju,Richard:2016eis,Lebed:2016hpi,Karliner:2017qhf}. It is now widely accepted that the spectrum of hadrons is not restricted to the excitations of ordinary mesons and baryons, and extends to multiquark configurations with hidden or open flavor.

Concerning the hidden-charm pentaquarks, there are many theoretical approaches: quark model approaches with different interactions~\cite{Yuan:2012wz,Santopinto:2016pkp,Wu:2017weo}, 
quark-cluster model~\cite{Takeuchi:2016ejt}, 
diquark models~\cite{Maiani:2015vwa,Li:2015gta,Ali:2016dkf,Lebed:2015tna,Zhu:2015bba,Zhu:2015bba}, and hadronic molecular picture~\cite{Wu:2010jy,Wu:2010vk,Garcia-Recio:2013gaa,Karliner:2015ina,Chen:2015loa,Roca:2015dva,He:2015cea,Meissner:2015mza,Chen:2015moa,Uchino:2015uha,Shimizu:2016rrd,Yamaguchi:2016ote,Shimizu:2017xrg}.   The observation channel consists of a proton and a $\Jpsi$, suggesting as quark content $c\bar c$ attached to light quarks. In the hadronic molecular description, the $P_c$ states couple to both open charm and hidden charm hadrons in channels such as $\Jpsi \,p$ or $\bar D\Sigma_c$, and hence whatever dynamical scheme is adopted at start, the detailed properties are sensitive to all surrounding thresholds and their interactions. 

Within the quark-model picture, the technical aspect of this competition between multiquark and hadron-hadron components deals with finding a reliable solution of the model in the continuum. In the literature on the quark model (including by some of us), the bound-state formalism  has often been used without caution for states in the continuum. To be more specific, if a crude trial wave function  gives for some multiquark configuration an energy $E=-100\,$MeV below the lowest threshold, it is  not too far from the exact solution within this model. On the other hand, an energy $E=+100\,$MeV above one of the thresholds might be  meaningless. Refining the trial wave  function  should lead to convergence towards the lowest threshold. Identifying a resonance within a given model requires dedicated methods. One of such methods is the technique of real scaling along the fall-apart coordinate, starting from a rich variational function that includes several types of clustering.

In Ref.~\cite{Hiyama:2005cf}, two of the present authors (E.H. and A.H.) studied the $\Theta^+$ state as a $udds\bar s$ five-body problem, taking explicitly into account the $K\,N$ scattering channel. This was, to our knowledge, the first application of real scaling to a quark model calculation. The present article is devoted to the hidden-charm pentaquarks $P_c$, treated as a $uudc\bar c$ system, with account for all possible open channels such as $\Lambda_c +D^*$, $\Sigma_c+D$, $\Sigma_c^*+D$, etc. This is a rather delicate and lengthy calculation, but rather rewarding: most states appear as building a mere discretization of the continuum, but there are some striking exceptions which can be identified as genuine resonances in the model. 

The paper is organized as follows. In Sec.~\ref{sec:model} we present Hamiltonian and the method. 
The results are displayed and discussed in Sec.~\ref{sec:resu}.
Finally we summarize our findings in Sec.~\ref{sec:sum}.

%========================
\section{Hamiltonian and method}\label{sec:model}
%========================

The Hamiltonian, which   corresponds to a  standard non-relativistic quark model,  is
given by
\begin{equation}\label{eq:H}
H=\sum_i (m_i + \frac{\vec p^2_i}{2m_i})-T_G - \frac{3}{16}\sum_{i<j} 
\lambda_i.\lambda_j\,V_{ij}(r_{ij}),
\end{equation}
where $m_i$ and $\vec p_i$ are the mass and momentum of
the $i^{\rm th}$ quark, $\lambda_i$ represents the  eight color-SU(3) operators for the $i^{\rm th}$ quark 
and $T_G$ is the kinetic energy of the center-of-mass
system. Here, we label the light quarks, $u$ and $d$, by $i=1$ to $3$, 
and heavy (charm) quarks, $c$ by $i=4$  and $\bar{c}$ by $i=5$. 

For the quark-quark interaction, we use the  potentials proposed by Semay and 
Silvestre-Brac \cite{Semay:1994ht,Silvestre}.
The functional form of the potential in the $r$-space is given by
\begin{multline}\label{eq:V}
V_{ij}(r)= -\frac{\kappa}{r} +\lambda r^p-\Lambda \\
{}+ \frac{2\pi\kappa'}{3m_im_j}
\frac{\exp(-r^2/r_0^2)}{\pi^{3/2}\,r^3_0}\,\vec\sigma_i.
\vec\sigma_j\,,
\end{multline}
where the smearing parameter $r_0$ is  a function of the quark masses, $r_0(m_i,m_j)=A(\frac{2m_im_j}{m_i+m_j})^{-B}$.
Two sets of parameter choices, AP1 and AL1, are employed in the current analysis and listed in Table I.
%%%%%%%%%%%%%%%%%%%%%%  Table I critical  %%%%%%% %
\begin{table*}[htb] \begin{center}
\caption{Parameters of AP1 and AL1 defined by Eq. (2.2).}
\begin{ruledtabular}
\begin{tabular}{cccccccccc}
\noalign{\vskip 0.1 true cm}
  & $p$  &$m_{u,d}$(GeV)  &$m_s$(GeV)  &$\Lambda$(GeV)  &$B$  &$A$(GeV$^{B-1}$)
  &$\kappa$  &$\kappa$'  &$\lambda$(GeV$^{5/3}$)   \\
 \noalign{\vskip 0.1 true cm} \hline
 \noalign{\vskip 0.15 true cm}
AP1  &$3/2$ &0.277 &0.553 
 &1.851    &0.3263 &1.5296  &0.5871 &1.8025  &0.3898  \\
 \noalign{\vskip 0.15 true cm}
AL1  &$1$ &0.315 &0.577  &1.836    &0.2204 &1.6553  &0.5069 &1.8609  &0.1653  \\
 \noalign{\vskip 0.15 true cm}
\end{tabular}
\end{ruledtabular}
\end{center}
\end{table*}
%%%%%%%%%%%%%%%%%%%%%%%%%%%%%%%%%%%%%%

Both sets reproduce rather well the masses of the ground states of heavy meson and baryon systems.
In Table~\ref{table:ET}, the calculated spectra using AP1 are listed.
The mass spectrum by AL1 is almost identical to that by AP1.

%%%%%%%%%%%%%%%%%%%%%%  Table I critical  %%%%%%% %
\begin{table}[htb] \begin{center}
\caption{The calculated masses (in MeV) of the heavy mesons and baryons entering the thresholds together with
the experimental values.}
\label{table:ET}
\begin{ruledtabular}
\begin{tabular}{ccrr}
\noalign{\vskip 0.1 true cm}
hadron  & $J^{P}$  &cal.\  &exp.\  \\
 \noalign{\vskip 0.1 true cm} \hline
 \noalign{\vskip 0.15 true cm}
 $\eta_c$ & $0^-$ &2984  & 2983 \\
$\Jpsi$ & $1^-$ &3103  & 3096 \\
$D$ & $0^-$ &1882  & 1869 \\
$D^*$ & $1^-$ &2033  & 2007 \\
$N$ & $1/2^+$ & 937  & 938 \\
$\Lambda_c$ & $1/2^+$ &2290  &2286 \\
$\Sigma_c$ & $1/2^+$ &2472  & 2455 \\
$\Sigma_c^*$ & $3/2^+$ &2545  & 2520 
\end{tabular}
\end{ruledtabular}
\end{center}
\end{table}
%%%%%%%%%%%%%%%%%%%%%%%%%%%%%%%%%%%%%%

\begin{figure*}[htb]
\centerline{
\includegraphics[scale=0.35]{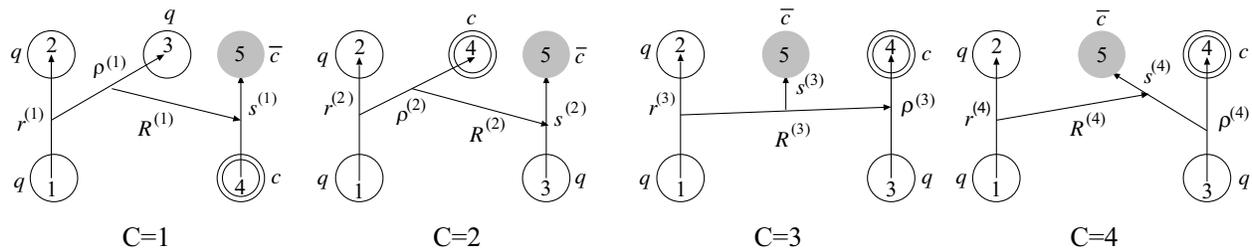}}
\caption{
Four sets of the Jacobi coordinate systems. The light ($u,d$) quarks, labeled by particle $1-3$,
are to be anti-symmetrized, while the particles 4 and 5 stand for $c$ and $\bar{c}$, respectively.
Scatterings of $qqq +c\bar{c}$ and $qqc+q\bar{c}$ are described in the coordinates $C=1$ and 2,
respectively.}
\label{fig:Jacobi}
\end{figure*}

The five-body  Schr\"{o}dinger equation
\begin{equation}\label{eq:SE}
(H-E) \Psi_{JM}=0\,
\end{equation}
corresponding to the Hamiltonian \eqref{eq:H} is solved by using
the Gaussian Expansion Method (GEM)
\cite{Kamimura1988,Hiyama2003} which  was successfully applied to various types of three- and four-body
systems \cite{Hiyama2004,Hiyama2009,Hiyama-hyper}. For a variant, see, e.g., \cite{Mitroy-RMP}. 
We describe the five-body wave function 
using the  four types of Jacobi coordinates  shown in Fig.~\ref{fig:Jacobi}.
Among them, $C=1$ and 2 are the configurations in which two color singlet clusters, such as
$\eta_c  N$, $\Jpsi  N$, $\Lambda_c  D$, $\Sigma_c  D$,  $\Sigma_c  D$  and so on
may fall apart along 
the inter-cluster coordinate, ${\vec R}^{(c)}$ ($C=1,2$).
Namely, for $C=1$, the color wave function is chosen as the product of
color-singlet $qqq$ plus color-singlet $c\bar{c}$, which corresponds to $\eta_c N$ and $\Jpsi N$ configurations.
For $C=2$, 
it is given as color-singlet $qqQ$ plus color-singlet $q\bar{c}$, which corresponds to $\Lambda_c D$, $\Sigma_c D$, and $\Sigma_c D$ configurations.
In contrast, the other two configurations $C=3$ and 4 do not describe color-singlet subsystems 
falling apart, and represent the five quarks as always connected by a confining interaction. 
In this sense, we call $C=3$ and 4 as the ``connected" (confining) configurations.

%%%%%%%%%%%%%%%%%%%%%%%%  Fig. 1  %%%%%%%%%%%%%%%%%%%%
%%%%%%%%%%%%%%%%%%%%%%%%%%%%%%%%%%%%%%%%%%%%%%

The full wave function $\Psi_{JM}$ is  written as a sum of components, each described in terms of
one of the four Jacobi coordinate systems, namely
\begin{multline}\label{eq:psi}
\Psi_{JM} =\sum_{C}{\cal A}_{123}\;
\xi_1^{(C)}\eta_T^{(C)} \Big[\chi^{(C)}_{S(s\bar{s}\sigma)} \times \Big[ \big[ [\phi_{nl}^{(C)}({\vec r}^{(C)})\\
          \varphi_{\nu\lambda}^{(C)}
          (\rhovec^{(C)})]_\Lambda 
       \psi_{NI}^{( C)}({\vec R}^{(C)}) \big]_{I'}
 \tilde\psi_{N'K}^{(C)}({\vec s}^{(C)}) \Big]_L\Big]_{J^P M} 
\end{multline} 
where $C$ specifies the set of Jacobi coordinates 
and $\xi_1^{(c)}$, $\eta_t^{(c)}$ and $\chi^{(c)}_{S(s\bar{s}\sigma)}$
represent the color, isospin and spin wave functions, respectively.
The orbital wave functions with appropriate orbital angular momenta are described by
$\phi^{(C)}$, $\varphi^{(C)}$, $\psi^{(C)}$ and $\tilde\psi^{(C)}$.
${\cal A}_{123}$ denotes the anti-symmetrization operator for the light quarks (1,2,3).
We consider the states with the total isospin $T=1/2$ and the total spin $S=1/2$ or 3/2.
In the present analysis, we take the total orbital angular momentum $L=0$ only.
Then the total spin-parity is  either $J^P=1/2^-$ or $3/2^-$.
According to the LHCb experiment, the observed pentaquark states may have $5/2^-$ or $ 3/2^+$.
However, in the present calculation, the energies of the states with total angular momentum larger than 3/2 or with positive parity are located at much higher masses.

The color-singlet wave function, $\xi_1^{(C)}$, for each Jacobi configuration  is chosen as
\begin{equation}\label{eq:color-1}
\begin{split}
\xi_1^{(1)} & =  [(123)_1(45)_1]_1\,, \\ 
\xi_1^{(2)} & =  [(124)_1 (35)_1]_1\,, \\
\xi_1^{(3)} & =  [(12)_{\bar{3}}(34)_{\bar{3}}]_3 5]_1\,,\\
\xi_1^{(4)} & =  [(12)_{\bar{3}}[(34)_{\bar{3}}5]_3]_1\,.
\end{split} 
\end{equation}
The spin and isospin wave functions are given by
\begin{eqnarray}
\label{eq:iso}
&& \eta_T^{(C)}= [\eta_{1/2}(i) \eta_{1/2}(j)]_t \eta_{1/2}(k)]_{T=1/2}, \\[2pt]
&&\begin{aligned}
\label{eq:spin}
 &\chi^{(1)}_{S(s\bar{s}\sigma)}(123,4,5) = [[(12)_s3]_\sigma (45)_{\bar{s}}]_S, \\ 
&\chi^{(2)}_{S(s\bar{s}\sigma)}(123,4,5) =  [[(12)_s4]_\sigma (35)_{\bar{s}}]_S, \\
&\chi^{(3)}_{S(s\bar{s}\sigma)}(123,4,5) = [[(12)_s (34)_{\bar{s}}]_\sigma 5]_S,  \\
&\chi^{(4)}_{S(s\bar{s}\sigma)}(123,4,5) = [(12)_s [(34)_{\bar{s}}] 5]_\sigma]_S.  
  \end{aligned}
\end{eqnarray}

The spatial wave functions for each channel $(C)$ are expanded by multi-range Gaussians multiplied by spherical harmonics as
\begin{equation}\label{eq:4gauss}
\begin{split}
&\phi_{nlm}({\vec r}) = N_{nl}\,r^l\:e^{-(r/r_n)^2}\: Y_{lm}({\hat {\vec r}})   \;, \\
&\varphi_{\nu \lambda \mu}(\rhovec)=N_{\nu \lambda}\,\rho^\lambda\:e^{-(\rho/\rho_\nu)^2}\:Y_{\lambda \mu}({\hat {\rhovec}})\;, \\ 
&\psi_{NIM}({\vec R}) =N_{NI}\,R^I\:e^{-(R/R_N)^2}\:Y_{IM}({\hat {\vec R}})\;, \\
&\psi_{N'KM_K}({\vec s}) = N_{N'K}\,s^K\:e^{-(s/s_{N'})^2}\:Y_{KM_K}({\hat {\vec s}})\;. 
\end{split}
\end{equation}
Here, it is important to choose the Gaussian ranges to
lie in geometric progression so that the basis functions are suitable for the 
descriptions of both short-range correlations and long-range asymptotic behavior  without introducing too many free parameters
 (see, for example,  Refs.~\cite{Kamimura1988,Kame89,Hiyama2003,
Hiya12FEW,Hiya12PTEP,Hiya12COLD,Ohtsubo2013}):
\begin{equation}\label{eq:r-rho-R-s}
\begin{aligned}
& r_n=r_1\, a^{n-1}\;\quad\quad \:          &&(n=1\ldots n_{\rm max})\;,  \\
&\rho_\nu=\rho_1\, b^{\nu-1}\;              &&(\nu=1\ldots\nu_{\rm max}) \;,\\ 
&R_N=R_1\, A^{N-1} \;                       &&(N=1\ldots N_{\rm max}) \;, \\
&s_{N'}=s_1\, A'^{N'-1} \;                  &&(N'=1\ldots N'_{\rm max}) \;.
\end{aligned}
\end{equation}

In Eqs.~\eqref{eq:4gauss} and  \eqref{eq:r-rho-R-s}, the channel index $(C)$ is omitted for simplicity, for example, $\vec{r^{(c)}}$ is replaced by
 $\vec{r}$.
The dimension of the basis of Gaussian wave functions, $n_{\rm max}$,
$\nu_{\rm max}$, $N'_{\rm max}$ for $C=1$ to  $4$ channels and 
$N_{\rm max}$ for $C=3$ and 4 channels is 5 or 6.
And there are  $N_{\rm max}=10$ Gaussian basis functions 
for $C=1$ and 2. Since these  are scattering channels, it is necessary to have many
basis functions.
Thus in diagonalizing the five-body Hamiltonian for $J^P =1/2^-$ and  $3/2^-$ states, we use about 40,000 basis functions, resulting in
the same number of eigenstates for each $J^{P}$.

It should be noted here that all the obtained eigenvalues are discrete, as the wave function for ${\vec R}^{(C=1,2)}$ is 
expanded on a finite basis of functions localized within $R\lesssim 2\,R_N$. 
Namely, even the continuum states corresponding to the baryon-meson scattering solutions
come out as discrete states.
Therefore when we look for a compact pentaquark state, appearing as a sharp resonance embedded in
the continuum, we need a method to distinguish the genuine resonances from the discretized scattering states.
Here we adopt the real-scaling (stabilization) method, often used for analyzing electron-atom and electron-molecule scattering~\cite{real-scal}, and already introduced in a previous quark-model calculation~\cite{Hiyama:2005cf}. 
In the present case, as we have explained in the definition of the Jacobi coordinate systems, the
continuum spectrum  arises only from the factors of the wave functions that depend on ${\vec R}^{(1)}$ or 
${\vec R}^{(2)}$.
The factors of  wave functions that depend on the coordinates other than ${\vec R}^{(1)}$ and ${\vec R}^{(2)}$ 
cannot have asymptotic states due to their colored configurations.
We therefore scale the basis functions for the expansion of the wave functions along
 ${\vec R}^{(1)}$ and ${\vec R}^{(2)}$.
To do this, 
we multiply all the range parameters simultaneously by a factor as $R_N \to \alpha R_N$.
Then any continuum state will fall off towards its threshold, while a compact resonance state should 
stay as it is not affected by the boundary at a large distance.

%=======================
\section{Results} \label{sec:resu}  
%=======================

\begin{figure*}[htb]
\centerline{
\includegraphics[scale=0.30]{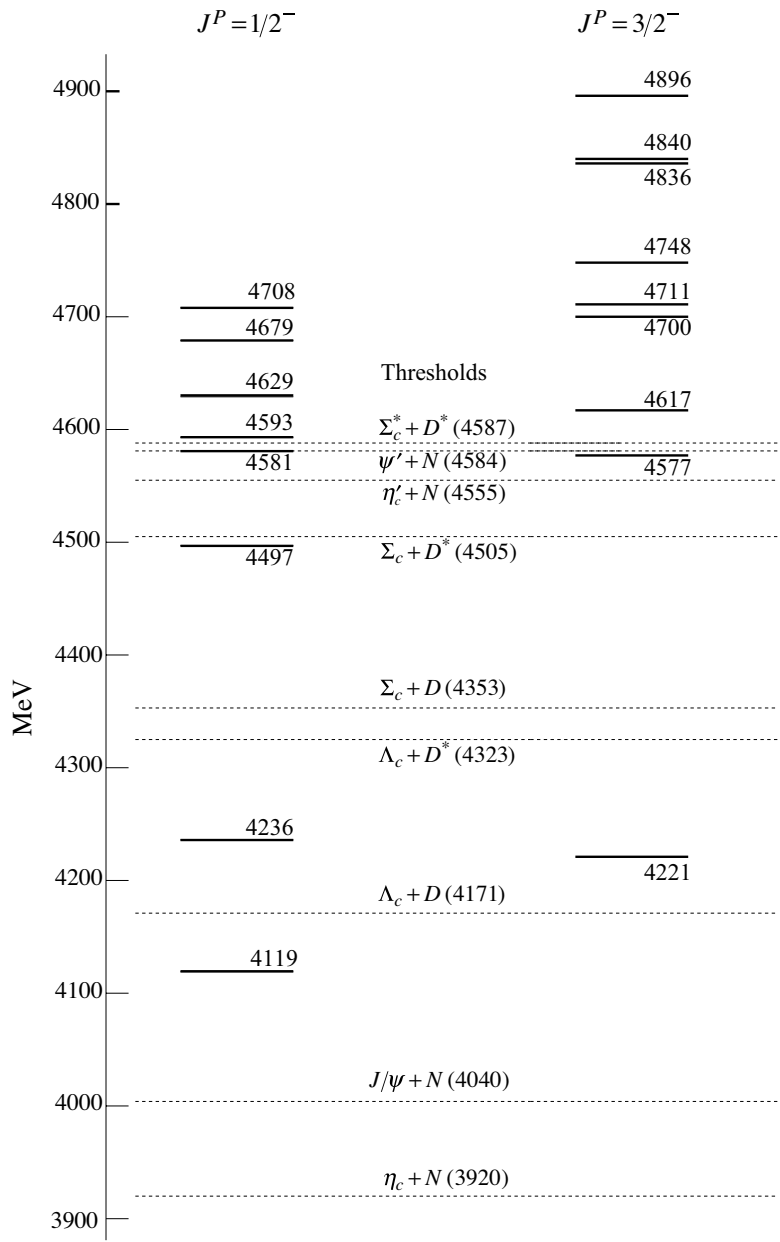}}
\caption{
The calculated energy spectra for $J=1/2^-$ and $3/2^-$ states
 using the connected  configurations of $C=3$ and 4.
The dashed lines are thresholds}
\label{fig:spectra3-4}
\end{figure*}

Let us start with the calculation without the contributions from ``continuum" (scattering) states.
This corresponds to a conventional estimate in the quark model
without coupling to fall-apart decaying channels.  
In our scheme, this can be done by including only 
the connected Jacobi coordinate systems, $C=3$ and 4 of Fig.~\ref{fig:Jacobi}.
Solving the five-body Schr\"odinger equation for spin-parity
$J^{P}=1/2^-$ and $3/2^-$ states, we obtain the masses shown in Fig.~\ref{fig:spectra3-4}.
We find that all the eigenvalues are found above the lowest meson-baryon threshold, $\eta_c N$ or $\Jpsi N$.
Nevertheless, as our wave function contains only the $C=3$ and 4 components, all the obtained states appear
as ``bound states'' without the contribution from scattering states.

%%%%%%%%%%%%%%%%%%%%%%%%  Fig. 2  %%%%%%%%%%%%%%%%%%%%
%%\onecolumngrid
%\begin{widetext}
%\twocolumngrid
%%\end{widetext}
%%%%%%%%%%%%%%%%%%%%%%%%%%%%%%%%%%%%%%%%%%%%%%

For the spin-parity $J^P=1/2^-$, the lowest state appears at 4119\,MeV, which is above the hidden charm threshold, 
$\eta_c N$ and $\Jpsi N$, and below the open charm one of 
$\Lambda_c D$. This result is compatible with, e.g., Ref.~\cite{Richard:2017una}, which uses the potential AL1. 
The second and third states appear at 4236\,MeV and 4497\,MeV, respectively.
These states lie around the LHCb pentaquarks (4380\,MeV and 4450\,MeV) 
though their spin-parity assignments
are different from those of the preferred ones ($3/2^-$ and $5/2^+$ respectively).
In addition, we find several states around and above 4500\,MeV.
For $J=3/2^-$,
the lowest state appears at 4221\,MeV, whose energy 
is higher than the lowest $J^P = 1/2^-$ state by about 100\,MeV. It is, again compatible with Ref.~\cite{Richard:2017una}.
It is followed by the second state at 4577\,MeV, which is a 
region of several open channel thresholds.  

For complete analysis, we need to include the scattering configurations
such as $qqq +c\bar{c}$ and $qqc+q\bar{c}$.
This can be achieved by including the $C=1$ and 2 configurations in our formulation.
Now,  the coupling of the scattering states 
may cause some (in fact many) of the ``bound states'' to disappear, 
melting away into the continuum spectrum.
This was already pointed out for the $\Theta^+$ pentaquark \cite{Hiyama:2005cf}. 
In the present case, the number of thresholds is significantly larger than that  
for the $\Theta^+$ system, 
and therefore the effects of the coupling is more pronounced.
For instance, we have nine thresholds opening within 700\,MeV 
from the lowest threshold in $J=1/2^-$ configurations, as indicated by the dashed lines 
in Fig.~\ref{fig:spectra3-4}.  

In order to investigate the nature of the bound states 
shown in Fig.~\ref{fig:spectra3-4}  
and their fate due to the coupling to scattering states, 
we investigate the behavior of the connected states 
coupled by a  scattering state one by one in 
the real scaling method.  
Namely, we  scale the range parameter $R_N$ of the Gaussian basis (see in \eqref{eq:4gauss})  as $R_N\to \alpha R_N$  for the Jacobi coordinates of $C=1$ or 2
i.e., $R_N = R_N^{(1)}$ and $R_N^{(2)}$. 
The eigenvalues corresponding to scattering states will fall down towards the threshold of
the scattering state as the scaling factor $\alpha$ increases.
On the other hand, resonance states will stay at a resonance energy independently
from the scaling parameter~$\alpha$. 

\begin{figure*}[htb]
\centerline{
\includegraphics[scale=0.4]{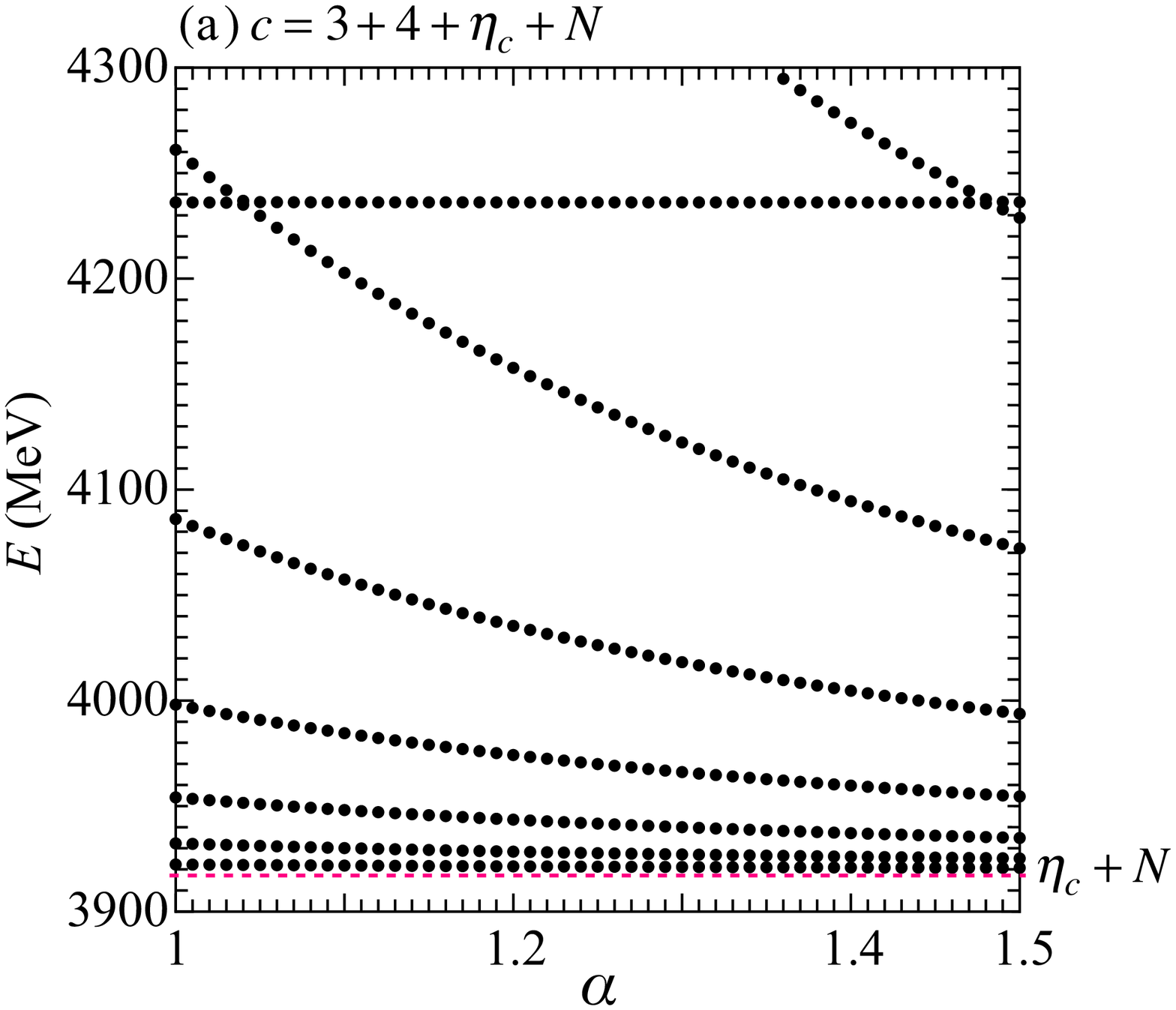}
\includegraphics[scale=0.4]{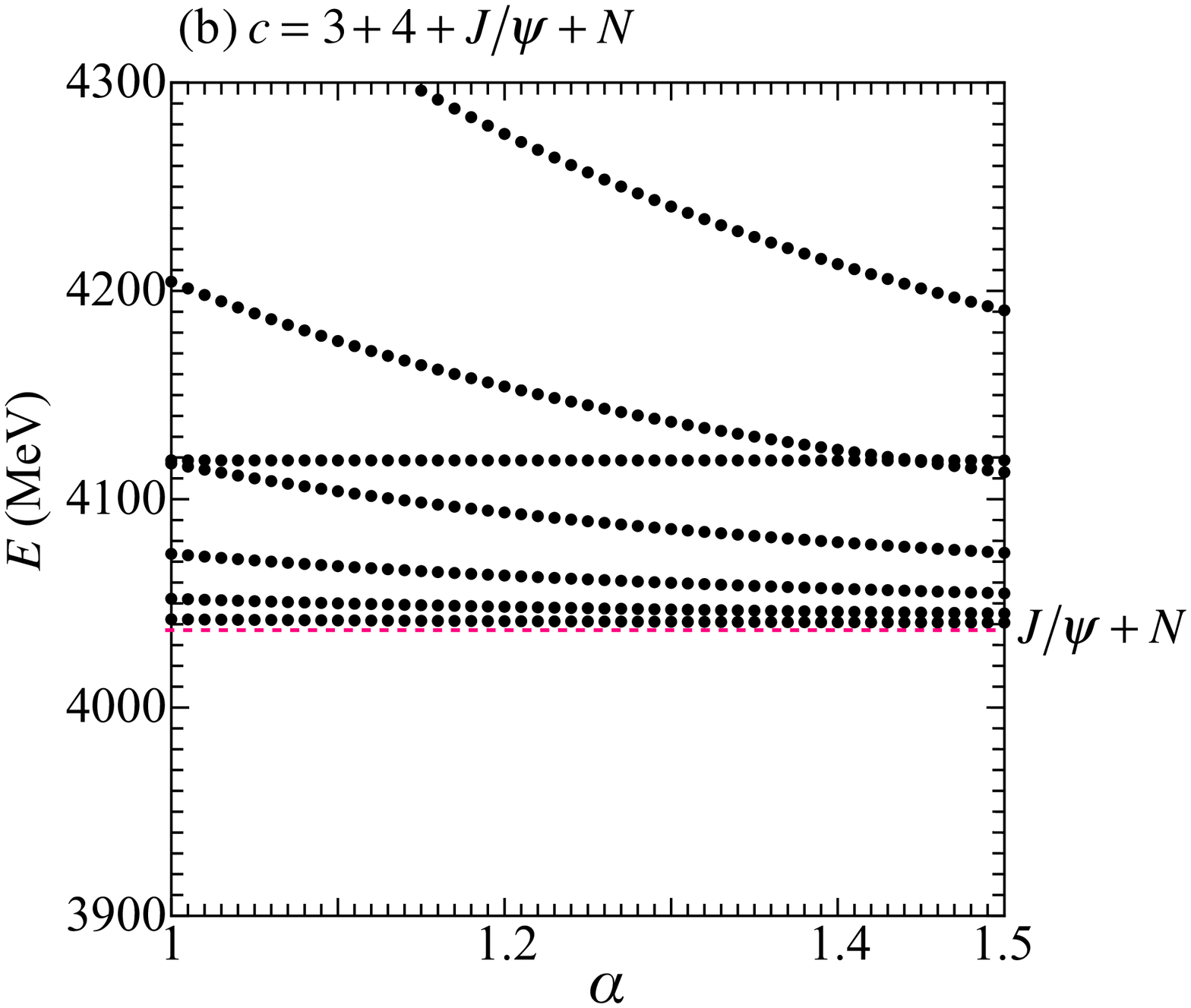}
}
\caption{
(color online)  The stabilization plots of the eigenenergies $E$
of the psuedostates for $J=1/2^-$ with the respect to the scaling factor $\alpha$ 
in the case of (a) only connected  configuration of $C=3+4$, and
$\eta_c +N$ configuration,
and of (b)  only connected configurations of $C=3+4$ and $\Jpsi +N$.
}
\label{fig:spectra-3-4-J}
\end{figure*}

To illustrate our method, 
let us study the two  ``connected'' states, the one at 4119\,MeV with $1/2^-$  and the one at 4236\,MeV with $3/2^-$.
We see that there are two open thresholds, $\Jpsi N$ and $\eta_c N$ below 
the state at 4119\,MeV, as shown in Fig.~\ref{fig:spectra3-4}.  
We have performed calculations for the following two cases,  
(a) $C=3$ and 4 (connected configurations)  
with only  $\eta_c N$ 
and
(b)  $C=3$ and 4 (connected configurations) 
with only $\Jpsi N$.
The coupling of the $\eta_c N$ or $\Jpsi N$ corresponds to the inclusion of 
$c=1$ of Fig.~\ref{fig:Jacobi} as it contains a configuration of $qqq$ and $c \bar c$ linked by $R_1$.  
The results are shown in Fig.~\ref{fig:spectra-3-4-J}(a) and \ref{fig:spectra-3-4-J}(b), respectively, 
where we demonstrate discrete  eigenenergies
as functions of $\alpha$.
There are about 30,000 basis functions, and thus about  30,000 eigenstates as discrete states.
For the real scaling method, we scale the coordinate 
$R^{(1)}$ for $C=1$ configuration by a scaling factor $1.0 < \alpha < 1.5$

%%%%%%%%%%%%%%%%%%%%%%  Fig. 3  %%%%%%%%%%%%%%%%%%%%
%%%%%%%%%%%%%%%%%%%%%%%%%%%%%%%%%%%%%%%%%%%%%%

As seen in Fig.~\ref{fig:spectra-3-4-J}(a),  we cannot find stable states  around 4119\,MeV:
all eigenstates around this mass move towards the $\eta_c N$ threshold when $\alpha$ increases.
This implies that the connected state at 4119\,MeV  strongly couples to  $\eta_c N$
and melts away into the scattering state in the complete 
five-body calculation.  

On the other hand, the connected state at 4236\,MeV stays stable.  
A level repulsion (avoided crossing) is seen 
at $\alpha=1.04$ and $\alpha=1.48$.
In the method of real scaling, the distance between the two levels at the repulsion point 
is related to the decay width.
In the present case, the distances are
small,  which indicates that the state is a very sharp resonance where
the connected state at 4236\,MeV couples only weakly with $\eta_c N$.  

To know the nature of the two connected states better, let us see
Fig.~\ref{fig:spectra-3-4-J}(b) which shows the result of the calculation with $C=3+4+J/\psi N$ configurations.
Now we see the opposite situation from Fig.~\ref{fig:spectra-3-4-J}(a). 
Namely, the connected state at 4119\,MeV stays stable, while the one at 4236\,MeV 
disappears.  

From these observations, we may interpret that the state at 4119\,MeV is dominated by 
$\eta_c$ and $N$ clusters,
and the one at 4236\,MeV by $J/ \psi$ and $N$ clusters.  
These clusters interact only weakly and cannot hold bound nor resonant states.  
The appearance of the bound states in Fig.~\ref{fig:spectra3-4} is due to the inclusion of only 
the connected diagrams, $C=3$ and 4 of Fig.~\ref{fig:Jacobi}.  
We have seen that 
the two lowest connected states, the one at 4119\,MeV and the one at 4236\,MeV disappear 
in the full five-body calculation, and that 
we do not have any resonant state 
in the energy region from 3900\,MeV to 4300\,MeV.

We have repeated the same procedure to analyze each connected state given in Fig.~\ref{fig:spectra3-4}
and  summarized the results 
in Table \ref{table:ED} (a) for $J^P = 1/2^-$ and 
in Table \ref{table:ED} (b) for $J^P = 3/2^-$, 
where the results are shown for the dominant scattering configurations which couple to each connected state.  
Table II (a) shows that most of the $J=1/2^-$ connected states 
in the energy region from $E=4119$\,MeV
to $E=4673$\,MeV have significant coupling to scattering states, 
and they do not  survive as resonances.
For instance, the state at 4497\,MeV which is close to
the observed $P_c$ couples strongly to
$\eta_c N$, $\Lambda_c D^*$ and $\Sigma_c D$ configurations.
In the present quark model Hamiltonian, there is not a sufficiently strong 
attractive force between these hadrons, 
and therefore, the bound states in the connected configurations 
do not survive as resonant states.  

\begin{figure*}[htb]
\centerline{
\includegraphics[scale=0.4]{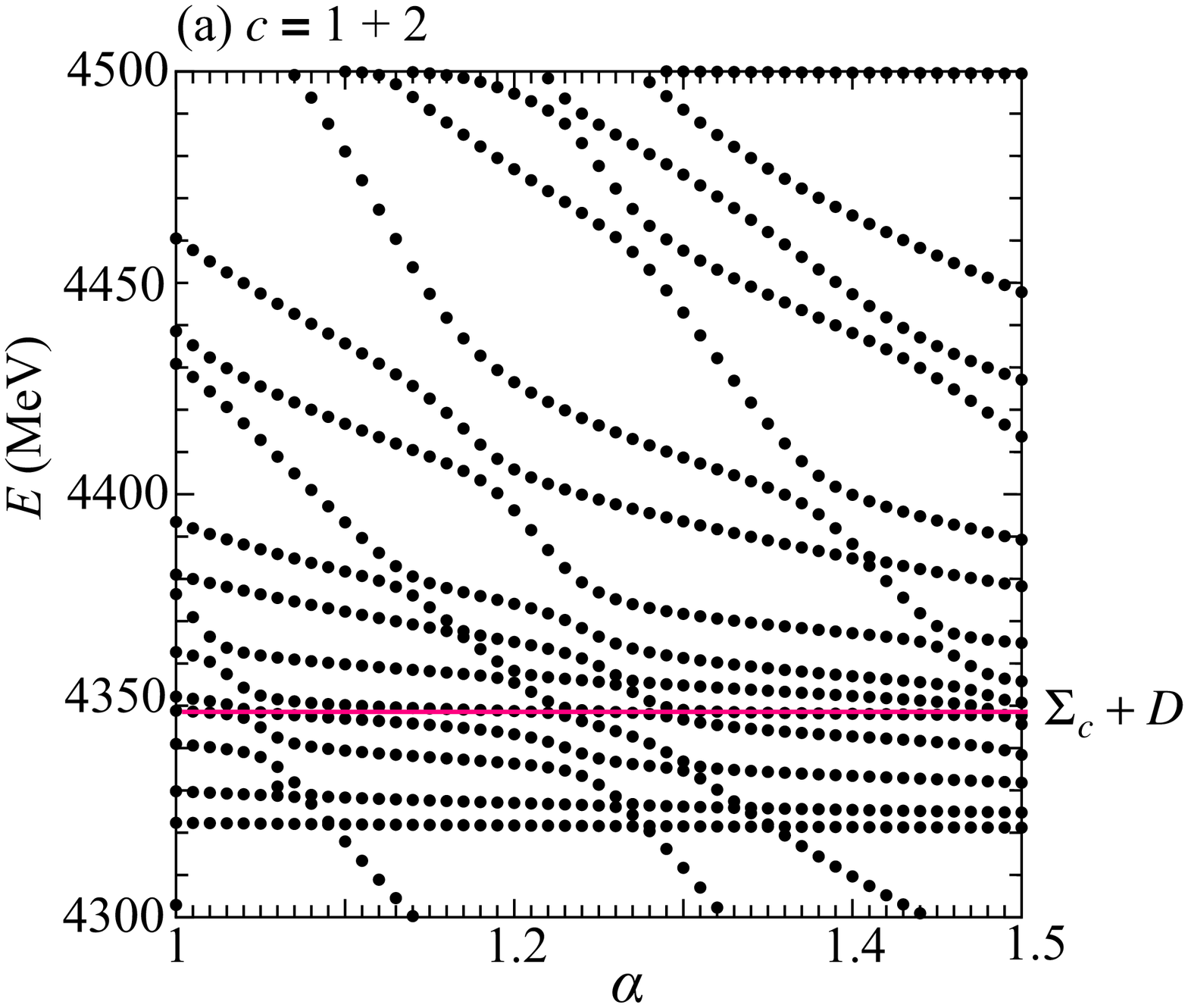}
\includegraphics[scale=0.4]{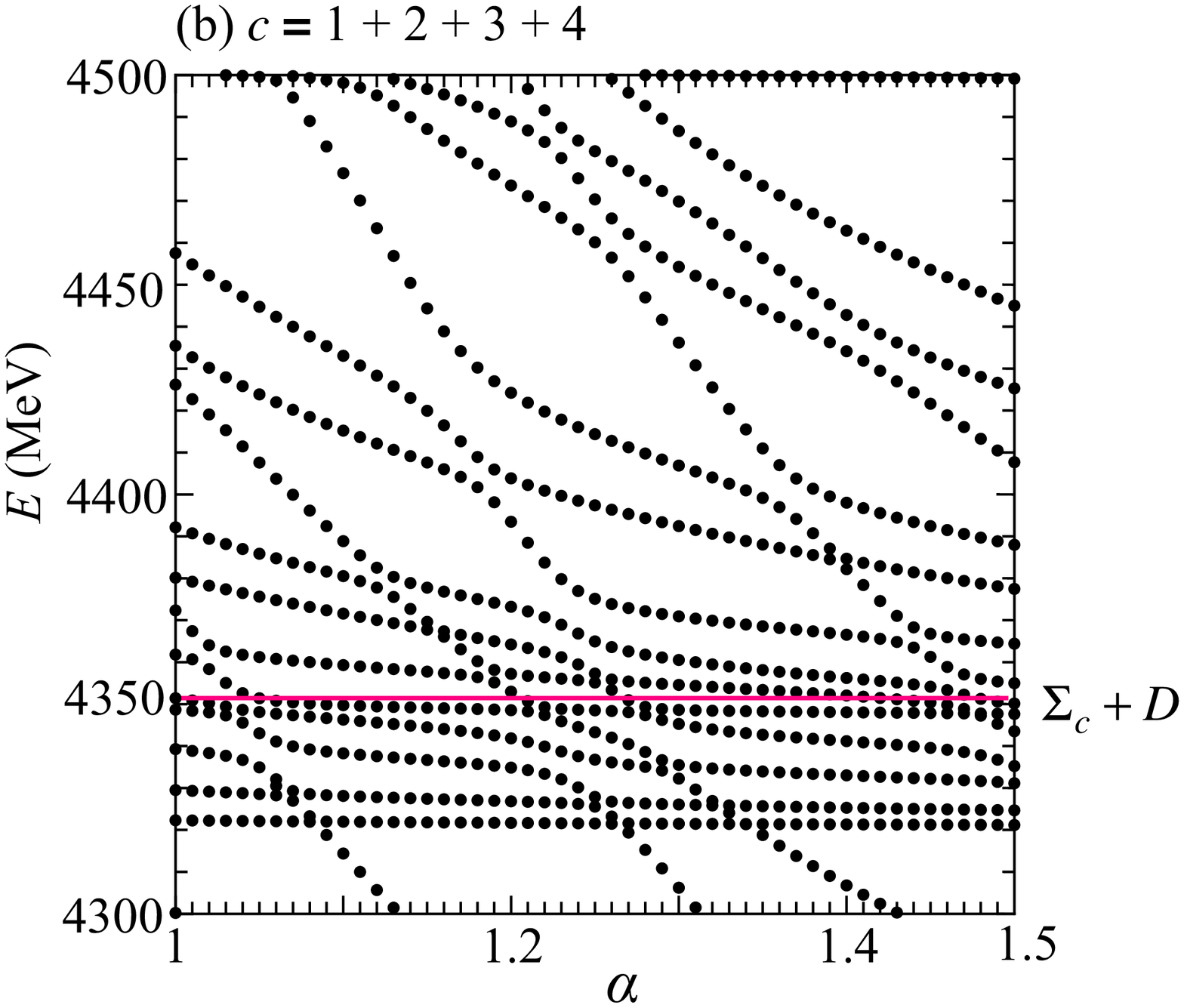}
}
\caption{
(color online)  The stabilization plots of the eigenenergies $E$
of the pseudostates with the respect to the scaling factor $\alpha$ 
in the case of (a) only scattering configurations of $C=1$ and 2,
and of (b)  full configurations of $C=1$ to $4$.
The Gaussian ranges $R_N$ for the coordinates ${\vec R_1}$ and ${\vec R_2}$ of
$C=1$ and 2 configurations are scaled as $R_N \rightarrow \alpha R_N$ 
with $1.0 $ to $1.5$. 
}
\label{fig:spectrap}
\end{figure*}

\begin{figure*}[htb]
\centerline{
\includegraphics[scale=0.4]{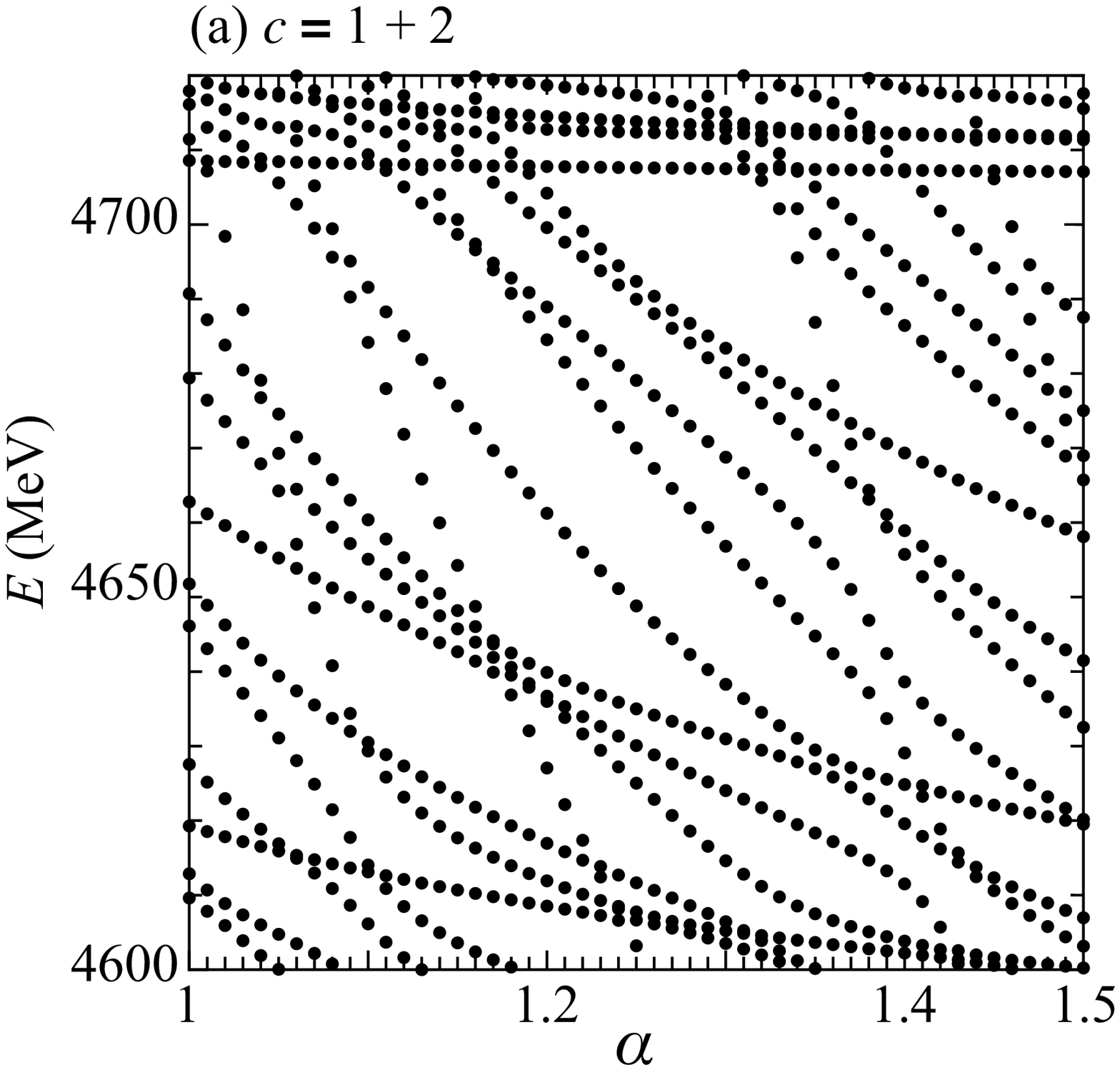}
\includegraphics[scale=0.4]{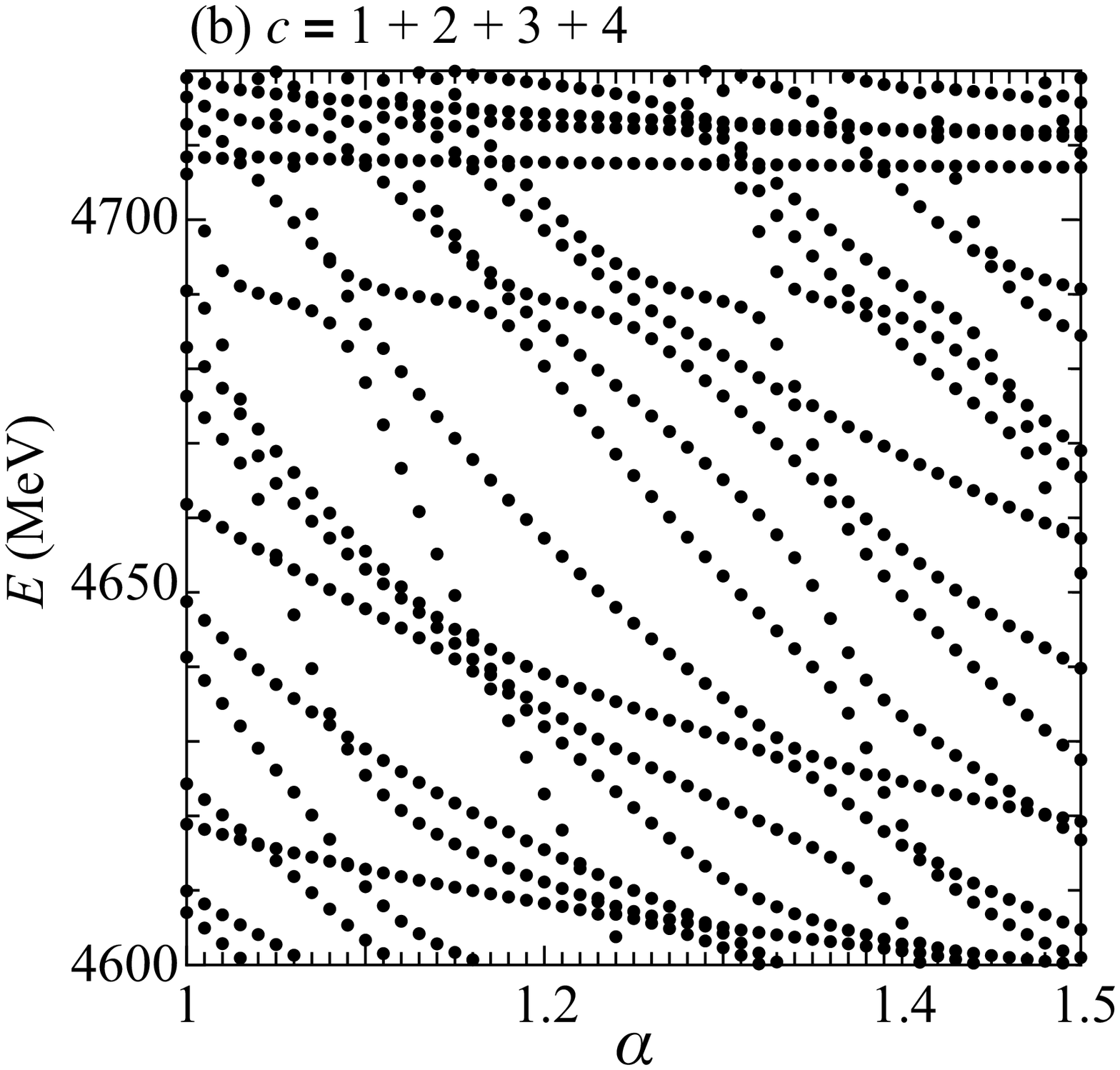}
}
\caption{
(color online)  The stabilization plots of the eigenenergies $E$
of the pseudostates with the respect to the scaling factor $\alpha$ 
in the case of (a) only scattering configurations of $C=1$ and 2,
and of (b)  full configurations of $C=1$ to $4$.
The Gaussian ranges $R_N$ for the coordinates ${\vec R_1}$ and ${\vec R_2}$ of
$C=1$ and 2 configurations are scaled as $R_N \rightarrow \alpha R_N$ 
with $1.0 $ to $1.5$. 
}
\label{fig:spectrapp}
\end{figure*}

There are, however, exceptions; 
the states at 4708\,MeV for $1/2^-$ and at 4896\,MeV for $3/2^-$.  
They do not have any dominant cluster structure, and likely corresponds
to a complicated five-quark configuration.
To see this point better, we show the stabilization plots in Figs.~\ref{fig:spectrap} 
for the case of $1/2^-$.
Figure~\ref{fig:spectrap}(a) shows the result when only the scattering configurations $C = 1$ and 2 are included, 
while (b) is for the full calculation.
As expected there is not resonance structure there.  
However, by including the ``seed" in connected configurations $C = 3$ and 4, 
we find a resonance structure at around 4690\,MeV as shown in Fig.~\ref{fig:spectrapp},  
which is identified with the Feshbach resonance with its seed as
the bound state at 4708\,MeV in the connected configurations.  
The same analysis for $3/2^-$ indicates that the resonance appears at around 4920\,MeV 
with the seed at 4896\,MeV.  

%%%%%%%%%%%%%%%%%%%%%%  Fig. 4  %%%%%%%%%%%%%%%%%%%%
%%%%%%%%%%%%%%%%%%%%%%%%%%%%%%%%%%%%%%%%%%%%%%

%%%%%%%%%%%%%%%%%%%%%%  Fig. 5  %%%%%%%%%%%%%%%%%%%%
%%%%%%%%%%%%%%%%%%%%%%%%%%%%%%%%%%%%%%%%%%%%%%

%%%%%%%%%%%%%%%%%%%%%%  Fig. 5  %%%%%%%%%%%%%%%%%%%%
\begin{figure}[htb]
\centerline{
\includegraphics[scale=0.4]{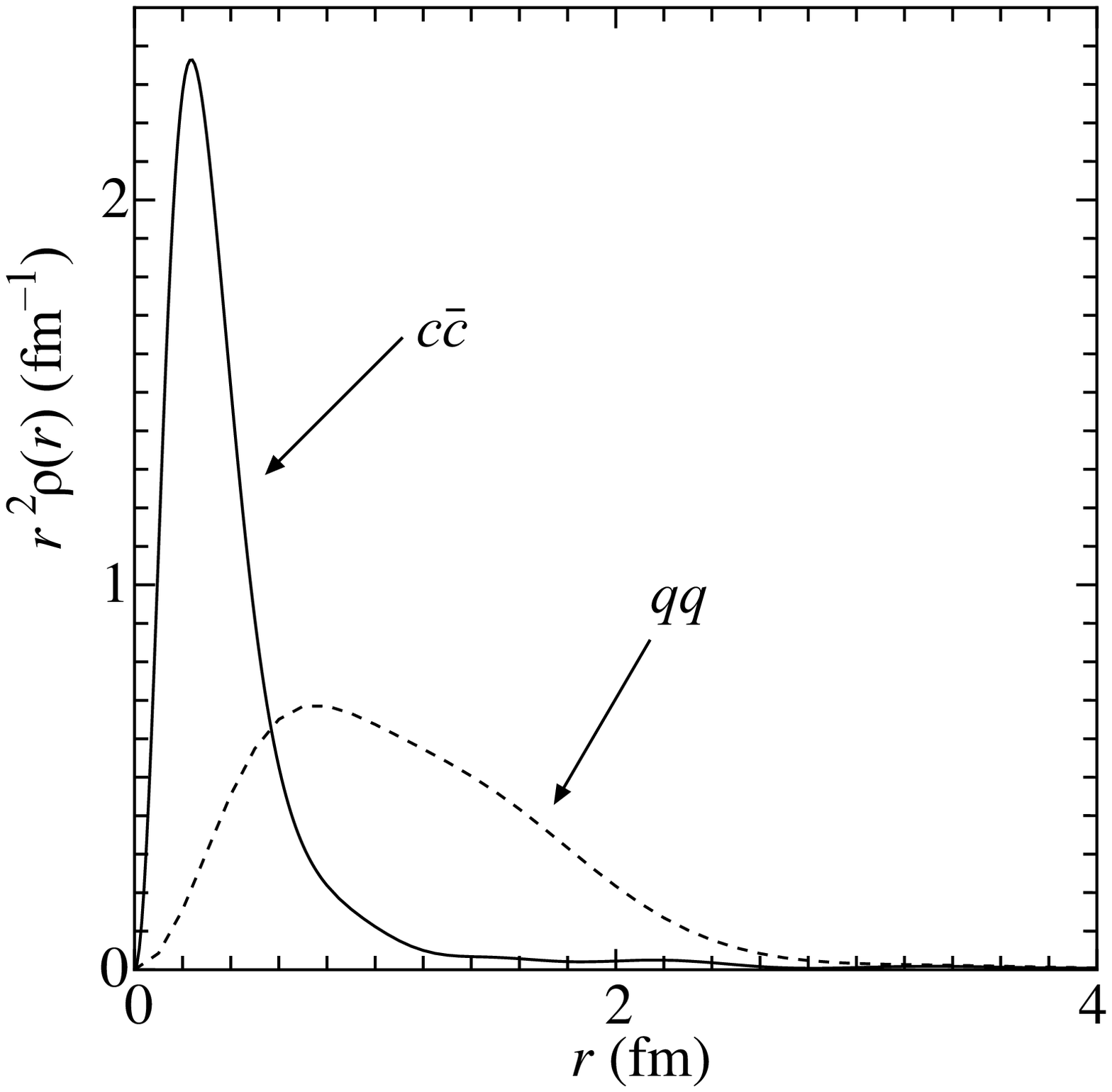}
}\caption{
The calculated $qq$ and $c\bar{c}$ correlation function for $J=1/2^-$ state with full configurations.
The solid line for $qq$ and the dashed line is for $c\bar{c}$.
}
\label{fig:spectrappp}
\end{figure}
%%%%%%%%%%%%%%%%%%%%%%%%%%%%%%%%%%%%%%%%%%%%%%

Now let us see how the above results depend on the choice of the Hamiltonian.
For this purpose,
we have also used the AL1 potential in  Ref.~\cite{Semay:1994ht,Silvestre}.
This potential also reproduces the ground states of the heavy mesons and
baryons.
We have  found that the results of our five-body calculations 
are essentially not modified, but with small changes in resonance masses, 
4690\,MeV for $J=1/2^-$ and 4920\,MeV  for $J=3/2^-$.

Finally, we would like to study the structure of  the resonant state found in the present work.  
We have calculated the two-body correlation function of $qq$ and of $c\bar{c}$ 
for the state at $E=4690$\,MeV for $J^P=1/2^-$.  
The correlation functions are defined as
\begin{equation}
\begin{split}
\rho_{qq}(r_1) &= \int |\Psi_{JM}|^2 d\vec{s}_1 d\vec{R}_1 d\rhovec_1 d\hat{\vec{r}}_1\\
\rho_{c \bar c}(s_1) &=  \int |\Psi_{JM}|^2 d\vec {r}_1 d \vec{R}_1 d\rhovec_1 d\hat{\vec{s}}_1\, ,
\end{split}
\end{equation}
where $\vec r_1$ and $\vec s_1$ are the relative distance between two light quarks $qq$ and $c \bar c$, 
as illustrated in Fig.~\ref{fig:Jacobi}, $C$ = 1.  
The integral was performed at $E=4689$\,MeV with the scaling factor $\alpha =1.05$,   
A consistency was checked by integrating over either $\vec r_1$ or $\vec s_1$, where 
we have verified that integrated normalization is $0.9999$. 
Figure~\ref{fig:spectrappp} shows the density distributions of $ r^2 \rho_{qq}$ and  $r^2 \rho_{c \bar c}$ 
as functions of the distance $r = \vec r_1 = \vec s_1$.  
We find that quarks in the pentaquark resonance distribute only within a compact region.  
The peak positions indicate that the light quarks extend about 0.8\,fm and  the charmed quarks are restricted to
about 0.2\,fm, which is consistent with the sizes of the nucleon $N$ and the charmonia $J/\psi$ or $\eta_c$.  

%===================
\section{Summary}\label{sec:sum}
%===================

Motivated by the observation of  pentaquark system of $P_c(4380)$ and
$P_c(4450)$, we solved the five-body scattering problem for the $J^P=1/2^-$ and $3/2^-$ state.with Gaussian expansion supplemented by real scaling.
Here we adopted non-relativistic quark model using AP1 potential proposed by Semay and 
Silvestre-Brac \cite{Semay:1994ht,Silvestre}.
The potential reproduces the experimental ground state energies of
heavy mesons and baryons entering the  open-channel thresholds relevant for the $P_c$ system. 

The main message is the clear possibility of distinguishing genuine resonances of the model from artifacts of the discretization.  This opens new perspectives for quark model calculations in the multiquark sector. 

In our calculation, based on a simple model of quark dynamics, two narrow states emerge, with a compact structure. They lie at  4690\,MeV for $J=1/2^-$ and at  4920\,MeV 
for  $J=3/2^-$, too high in mass to be identified with any  LHCb pentaquark. However, an improved quark model could probably lead to a better phenomenology. In particular, the model \eqref{eq:H} is probed only for color singlet and antitriplet: the value of the potential in the sextet and octet color states can perhaps be modified, for instance by introducing three- or four-body forces.  

We believe that our method, namely a Gaussian expansion of the wave function supplemented by real scaling,  can be used to more sophisticated models applied to the hidden-charm pentaquarks and to other multiquark configurations.

% We were not able to find any sharp resonant state at
% observed energy region. On the other hand, we found a resonant state at higher energy region, 4690\,MeV for $J=1/2^-$ and at  4920\,MeV 
% for  $J=3/2^-$ states, respectively,  which have
% penta-quark compacting structure.
% However, they reported the two states at LHCb experiment, which is inconsistent with
% our work.
% Here it should be noted that there is no interaction between $(3q)$ and $(c\bar{c})$ in $C=1$ configuration
% and $(qqc)$ and $(q\bar{c})$  in $C=2$ configuration in Fig.~\ref{fig:Jacobi}.
% Then, it is difficult to describe meson-bayon structure-like resonant states 
% in our present work.
% Therefore, from our results, it is possible to be indicated
%  that the observed states at LHCb are 
% meson-baryon structure like resonant states.  

%%%%%%%%%%%%%%%%%%%%%%  Table II  %%%%%%% %
\begin{table} [!thb] \begin{center}
\caption{Dominant hadron-hadron component for the various bound states in Fig.~\ref{fig:spectra3-4} structure of each energy taking only connected configurations
for (a)$J=1/2^-$ and $J=3/2^-$ states. Since the states at $E=4836$ and 4840 are close to each other,
their hadron-hadron components have been merged. 
The states at $E=4708$\,MeV and $E=4896$\,MeV have no significant overlap with
the  open channels shown in Fig.~\ref{fig:spectra3-4}. }
\label{table:ED}
\begin{ruledtabular}
\begin{tabular}{ccc}
\noalign{\vskip 0.1 true cm}
(a) $J=1/2^-$  &energy(MeV)   & configuration  \\
 \noalign{\vskip 0.1 true cm} \hline
 \noalign{\vskip 0.15 true cm}
& $4119$ &$\eta_c +N$ \\
&$4236$ &$J/\psi +N$, $\Lambda_c +D$ \\
&$4497$ &$\eta_c +N$,$\Lambda_c+D^*$, $\Sigma_c+D$ \\
&$4581$   &$J/\psi +N$ \\
&$4593$   &$\Lambda_c+D$ \\
&$4629$ &$\psi' +N$ \\
&$4679$ & $\Sigma_c +D^*$ \\
&$4708$ & - \\
\noalign{\vskip 0.15 true cm}
\hline
\noalign{\vskip 0.1 true cm}
(b) $J=3/2^-$  &energy(MeV)   & configuration  \\
 \noalign{\vskip 0.1 true cm} \hline
 \noalign{\vskip 0.15 true cm}
& $4221$ &$J/\psi +N$ \\
&$4577$ &$\Lambda_c+D^*, \Sigma_c*D^*, \Sigma_c^*+D^*$,$J/\psi +N$ \\
&$4617$ &$\Sigma_c^* +D$,$\Sigma_c^*+D^*$ \\
&$4700$   &$\Sigma_c^*+D$,$\Sigma_c^*+D^*$ \\
&$4711$   &$\Sigma_c+D^*$ \\
&$4748$   &$\Sigma_c^*+D$ \\
&$4836$ &$\Sigma_c^*+D^*$ or $\Sigma_c^* +D$\\
&$4840$ &$\Sigma_c^*+D^*$  or   $\Sigma_c^* +D$ \\
&$4896$ &- 
\end{tabular}
\end{ruledtabular}
\end{center}
\end{table}
%%%%%%%%%%%%%%%%%%%%%%%%%%%%%%%%%%%%%%

%%%%%%%%%%%%%%%%%%%%%%%%%%%%%
\begin{acknowledgments}
%%%%%%%%%%%%%%%%%%%%%%%%%%%%%
The authors thank Prof.~S.~Takeuchi and Prof.~M.~Takizawa for valuable discussion.
This work was supported by a
JSPS-Japan--CNRS-France Joint Research Project
and by the Grant-in-Aid for Scientific Research (No. 25247036 and JP26400273(C))
from the Japan Society for the Promotion of Science.

\end{acknowledgments}

\end{document}